\documentclass[11pt,a4paper]{article}
\usepackage{epsf,epsfig,amsfonts,amsgen,amsmath,amstext,amsbsy,amsopn,amsthm}
\usepackage{amssymb}
\usepackage{cite}
\usepackage{enumitem}
\theoremstyle{definition}

\begin{document}
\title{\bf {\Large New Quantum MDS codes constructed  from Constacyclic codes}}
\date{}

\author{Liangdong Lu$^{a,b,\dag}$, Wenping Ma$^{a}$,  Ruihu Li$^{b}$,   Yuena Ma$^{b}$, Luobin Guo$^{b}$,\\
 a.  School of Telecommunications Engineering,  Xidian University,  Xi'an, Shaanxi\\ 710051,\ China, (email:
$^{\dag}$ kelinglv@163.com, )\\
b. College of  Science,  Air Force Engineering University,  Xi'an,
Shaanxi 710051, China}

\maketitle

\begin{abstract} Quantum maximum-distance-separable (MDS) codes are an important
class of quantum codes. In this paper, using constacyclic codes and
Hermitain construction, we construct some new quantum MDS codes of
the form $q=2am+t$, $n=\frac{q^{2}+1}{a}$. Most of these quantum MDS
codes are new in the sense that their parameters are not covered be
the codes available in the literature.

\medskip

\noindent {\bf Index terms:} quantum MDS code, Hermitian
construction, cyclotomic coset,
 constacyclic code.
\end{abstract}

\section{\label{sec:level1} Introduction\protect}

Quantum error-correcting codes play an important role in quantum
information processing and quantum computation. As in classical
coding theory, one of the principal problems in quantum theory is to
construct quantum codes with the best possible minimum distance.
Calderbank et al.in \cite{cal1,cal2} found that one can construct
binary quantum codes from classical self-orthogonal codes over
$F_{2}$ or $F_{4}$ with respect to certain inner product. Nonbinary
case was generalized in \cite{Ash, Kai}. A lot of  quantum codes
have been constructed by using classical error-correcting codes with
Euclidean or Hermitian self-orthogonality\cite{Chen,Li1,Li2}.

Let $q$ be a prime power. A $q$-ary quantum code of length $n$ with
size $q^{k}$ is a $q^{k}$-dimensional subspace of the
$q^{n}$-dimensional Hilbert space. We use  $[[n,k,d]]_{q}$ to denote
a $q$-ary quantum code of length $n$ and minimum distance $d$, which
can detect up to $d-1$ quantum errors and correct up to
$\lfloor\frac{d-1}{2}\rfloor$ quantum errors. It is known that the
parameter of an  $[[n,k,d]]_{q}$ quantum code must satidfy the
quantum Singleton bound: $2d\leq n-k+2$\cite{Kai,Knill}. A quantum
code achieving this quantum Singleton bound is called a quantum
maximum-distance-separable (MDS) code. Quantum error-correcting
codes can protect quantum information, so that one can engineer more
reliable quantum communication schemes and quantum computers.
Construction good quantum error-correcting codes become a central
topic for quantum codes in recent years. Several quantum MDS codes
have been obtained by the Hermitian construction, which is one of
the most frequently used construction methods.

{\bf Theorem 1.1 \cite{Kai}.} (Hermitian Construction) If
$\mathcal{C}$ is an $[n,k,d]_{q^{2}}$ linear code such that
$\mathcal{C}^{\perp h}$$\subseteq \mathcal{C}$, where
$\mathcal{C}^{\perp h}$ denotes the Hermitian dual code of
$\mathcal{C}$, then there exists an $[[n,2k-n,\geq d]]_{q}$ quantum
code.

Using this Hermitian construction, many quantum MDS codes have been
constructed
\cite{Chen,He,Jin,Kai1,Kai2,Wang,Zhang,ZhangT1,ZhangT2,Taneja,shi,Qian}.
 However, it is not easy
task to construct quantum MDS codes with relatively large minimum
distance. Besides some special code length, most of known $q$-ary
quantum MDS codes have minimum distance less than or equal to
$\frac{q}{2}+1$. In this paper, we construct several families of
quantum MDS codes with length $n=\frac{q^{2}+1}{a}$ from classical
constacyclic codes by Hermitian construction. From \cite{Chen}, Chen
points that dual-containing constacylic codes over $F_{q^{2}}$ exist
only when the order $r$ is a divisor of $q+1$. Let $r=q+1$, we
construct quantum MDS codes from constacyclic codes.
 More precisely, Our main contribution on new
$q$-ary quantum MDS codes is as follows:

(1)  $$[[\frac{q^{2}+1}{13},\frac{q^{2}+1}{13}-2d+2,d]]$$

 where $q=26m+5$ is an odd prime power, and $2\leq d \leq 10m+2$ is even;

 where $q=26m+21$ is an odd prime power, and $2\leq d \leq 10m+8$ is even.

 (2)  $$[[\frac{q^{2}+1}{17},\frac{q^{2}+1}{17}-2d+2,d]]$$

 where $q=34m+13$ is an odd prime power, and $2\leq d \leq 10m+4$ is even;

 where $q=34m+21$ is an odd prime power, and $2\leq d \leq 10m+6$ is even.

 (3)  $$[[\frac{q^{2}+1}{25},\frac{q^{2}+1}{25}-2d+2,d]]$$

 where $q=50m+7$ is an odd prime power, and $2\leq d \leq 14m+2$ is even ;

 where $q=50m+43$ is an odd prime power, and $2\leq d \leq 14m+12$ is even.

 (4)  $$[[\frac{q^{2}+1}{29},\frac{q^{2}+1}{29}-2d+2,d]]$$

 where $q=58m+17$ is an odd prime power, and $2\leq d \leq 14m+4$ is even;

 where $q=58m+41$ is an odd prime power, and $2\leq d \leq 14m+10$ is even.

  (5)  $$[[\frac{q^{2}+1}{37},\frac{q^{2}+1}{37}-2d+2,d]]$$

 where $q=74m+31$ is an odd prime power, and $2\leq d \leq 14m+6$ is even;

 where $q=74m+43$ is an odd prime power, and $2\leq d \leq 14m+8$ is even.

  (6)  $$[[\frac{q^{2}+1}{41},\frac{q^{2}+1}{41}-2d+2,d]]$$

 where $q=82m+9$ is an odd prime power, and $2\leq d \leq 18m+2$ is even;

 where $q=82m+73$ is an odd prime power, and $2\leq d \leq 18m+16$ is even.

In construction (1), we obtain some quantum MDS codes with the
minimal distance larger than $\frac{q}{2}+1$. Comparing the
parameters with all known quantum MDS codes, we find that these
quantum MDS  codes are new in the sense that their parameters are
not covered by the codes available in the literature.

The paper is organized as follows. In Section 2, basic notations and
results about quantum codes and constacyclic codes are provided. In
Section 3, necessary and sufficient conditions for the existence of
dual-containing constacyclic codes are obtained. Then six classes of
quantum MDS codes are constructed through constacyclic codes.

\section{\bf Preliminaries}
In this section, we review some basic results on constacyclic codes,
BCH codes and QECCs for the purpose of this paper. For details on
BCH codes and constacyclic codes can be found in standard textbook
on coding theory \cite{Macwilliams,Huffman}, and
 for QECCs please see Refs. \cite{cal1,cal2,Ash,Kai,Li2}.

Let $p$ be a prime number and $q$ a power of $p$, ie., $q=p^{l}$ for
some $l>0$. $F_{q^{2}}$ denotes the finite field with $q^{2}$
elements. For any $\alpha \in F_{q^{2}}$, the conjugation of
$\alpha$ is denoted by $\overline{\alpha}=\alpha^{q}$. Given two
vectors $\mathbf{x}=(x_{1},x_{2},\cdots,x_{n})$ and
$\mathbf{y}=(y_{1},y_{2},\cdots,y_{n})\in F_{q^{2}}^{n}$, their
Hermitian inner product is defined as
$(\mathbf{x},\mathbf{y})_{h}=\sum \overline{x_{i}}y_{i}=\overline{x_{1}}y_{1}+\overline{x_{2}}y_{2}+\cdots+\overline{x_{n}}y_{n}.$
For a linear code $\mathcal{C}$ over $F_{q^{2}}$ of length $n$, the
Hermitian dual code $\mathcal{C}^{\bot _{h}}$ is defined as
 $\mathcal{C}^{\bot _{h}}=\{x\in  F_{q^{2}}^{n} | (x,y)_{h}=0, \forall  y $$\in \mathcal{C}\}$
If $\mathcal{C} \subseteq \mathcal{C}^{\bot _{h}}$, then
$\mathcal{C}$ is called a Hermitian dual containing code, and
$\mathcal{C}^{\bot _{h}}$ is called a Hermitian self-orthogonal
code.

We now recall some results about classical constacyclic codes,
negacyclic code and cyclic codes. For any
vector $(c_{0},c_{1},\cdots,c_{n-1}) $
$\in F_{q^{2}}^{n}$, a
$q^{2}$-ary linear code $\mathcal{C}$ of length $n$ is called $\eta$-constacyclic
if it is invariant
under the $\eta$-constacyclic shift of $F_{q^{2}}^{n}$:
$$
(c_{0}, c_{1}, \cdots , c_{n-1}) \rightarrow (\eta c_{n-1},
c_{0},\cdots, c_{n-2}),$$ where $\eta$ is a nonzero element of
$F_{q^{2}}$. Moreover, if
$\eta=1$, then $\mathcal{C}$ is called a cyclic code ; and if $\eta=-1$,
 then $\mathcal{C}$ is  called a negacyclic code.

 For a constacyclic code $\mathcal{C}$, each codeword $c =
(c_{0}, c_{1}, \cdots, c_{n-1})$ is customarily represented in its
polynomial form: $c(x) = c_{0} + c_{1}x + \cdots + c_{n-1}x_{n-1},$
and the code $\mathcal{C}$ is in turn identified with the set of all
polynomial representations of its codewords. The proper context for
studying  constacyclic codes is the residue class ring
$\mathcal{R}_{n}=\mathbb{F}_{q}[x]/(x^{n}-\eta)$. $xc(x)$ corresponds
to a constacyclic shift of $c(x)$ in the ring $\mathcal{R}_{n}$. As we
all know, a linear code $\mathcal{C}$ of length $n$ over $F_{q^{2}}$
is constacyclic if and only if C is an ideal of the quotient ring
$\mathcal{R}_{n}=\mathbb{F}_{q}[x]/(x^{n}-\eta)$. It follows that
$\mathcal{C}$ is generated by monic factors of $(x^{n}-\eta)$, i.e.,
$\mathcal{C}=\langle f(x) \rangle$ and $f(x)|(x^{n}-\eta)$. The $f(x)$
is called the generator polynomial of $\mathcal{C}_{n}$.

Let $\eta \in F_{q^{2}}$ be a primitive $r$th root of unity. Let
 $gcd(n,q)=1$, then there exists a primitive $rn$-th root
of unity $\omega$ in some extension field field of $F_{q^{2}}$ such that
$\omega^{n}=\eta$. Hence, $x^{n}-\eta =\prod ^{n-1}
_{i=0} (x-\omega^{1+ir})$.
Let $\Omega=\{1+ir|0\leq i \leq n-1\}$. For each $j\in \Omega$, let
$C_{j}$ be the $q^{2}$-cyclotomic coset modulo $rn$ containing $j$.
Let $\mathcal{C}$ be an $\eta$-constacyclic code of length $n$ over $F_{q^{2}}$
with generator polynomial $g(x)$. The set  $T=\{j\in\Omega|g(\omega^{j})=0\}$ is
called the defining set of  $\mathcal{C}$.
Let $s$ be an integer with $0\leq s < rn$, the
$q^{2}$-cyclotomic coset modulo $rn$ that contains $s$ is defined by
the set $C_{s}=\{s, sq^{2}, sq^{2\cdot 2}, \cdots, sq^{2(k-1)} \}$
(mod $rn$), where $k$ is the smallest positive integer such that
$xq^{2k}$ $\equiv x$ (mod $rn$).
We can see that the defining set $T$ is a union of some
$q^{2}$-cyclotomic cosets module $rn$ and $dim(\mathcal{C}) =
n-|T|$.

{\bf Lemma 2.1 \cite{Kai2}}. Let $\mathcal{C}$ be a $q^{2}$-ary
constacyclic code of length $n$ with defining set $T$. Then
$\mathcal{C}$ contains its Hermitian dual code if and only if $T
\bigcap T^{-q}=\emptyset$, where $T^{-q}$ denotes the set
$T^{-q}=\{-qz($mod  $rn)| z\in T\}$.

Let $\mathcal {C}$ be a constacyclic code with a defining set $T =
\bigcup \limits_{s \in S} C_{s}$. Denoting $T^{-q}=\{rn-qs | s\in T
\}$, then we can deduce that the  defining set of $\mathcal
{C}$$^{\bot _{h}}$ is $T^{\perp _{h}} =$$ \mathbb{Z}_{n}
$$\backslash T^{-q}$, see Ref. \cite{Kai2}.

Since there is a striking similarity between cyclic codes and
constacyclic code, we give a correspondence defining of skew aymmetric
and skew asymmetric as follows.

 A cyclotomic coset $C_{s}$ is {\it skew symmetric } if $rn-qs$ mod $rn\in
 C_{s}$; and otherwise is skew asymmetric otherwise. {\it  Skew asymmetric
 cosets}
$C_{s}$ and $C_{rn-qs}$ come in pair, we use $(C_{s},C_{rn-qs})$ to
denote such a pair.

The following results on $q^{2}$-cyclotomic  cosets,
dual containing BCH codes are bases of our discussion.

{\bf Lemma 2.2 \cite{Li4,Yang}.} Let $r$ be a positive divisor of
$q+1$  and $\eta\in F^{*}_{q^{2}}$ be of order $r$. Let gcd $(q, n)
= 1$, ord$_{rn}$$(q^{2}) = m$, $0 \leq x, y$, $z \leq n-1$.

(1) $C_{x}$ is skew symmetric if and only if there is a $t\leq \lfloor\frac{m}{2}\rfloor$
 such that $x \equiv xq^{2t+1}$(mod rn).

(2) If $C_{y}\neq C_{z}$, $(C_{y}, C_{z})$ form a skew asymmetric pair if and only if there is a
$t\leq \lfloor\frac{m}{2}\rfloor$
such that $y \equiv zq^{2t+1}$ (mod n) or $z \equiv yq^{2t+1}$(mod rn).

Thus one has the following lemma.

{\bf Lemma 2.3 \cite{Li4,Kai2}.} Let $r$ be a positive divisor of
$q+1$ and $\eta\in F^{*}_{q^{2}}$ be of order $r$. Let $\mathcal{C}$
be a $\eta$-constacyclic code of length $n$ over $F_{q^{2}}$ with
defining set $T$, then $\mathcal {C}$$^{\perp
_{h}}$$\subseteq\mathcal{C}$ if and only if one of the following
holds:

 (1) $T \cap$$T^{-q}=\emptyset$, where $T^{-q}=\{rn-qs \mid s\in
T\}$.

 (2) If $i,j,k\in T$, then
 $C_{i}$ is not a skew asymmetric coset and
($C_{j}$, $C_{k}$) is not  a skew asymmetric cosets pair.

{\bf Theorem 2.4 \cite{Yang}.} (The BCH bound for Constacyclic
Codes) Let $\mathcal{C}$ be an $\eta$-constacyclic code of length
$n$ over
 $F_{q^{2}}$, where $\eta$ is a primitive $r$th root of unity. Let $\omega$
 be a primitive $rn$-th root of unity in an extension field of $F_{q^{2}}$ such that
 $\omega^{n}=\eta$. Assume the generator polynomial of $\mathcal{C}$ has roots that include the set $\{\omega^{1+ri}|i_{1}\leq i \leq i_{1}+d-2\}$. Then the minimum distance of $\mathcal{C}$ is at least $d$.

\section{New  Quantum MDS Codes}

Let $q$ be an odd prime power with the form $2am+t$ and $a|t^{2}+1$,
where $a$ is an odd and $m$ is a positive integer. Then $a$ is a
divisor of $q^{2}+1$. Let $n=\frac{q^{2}+1}{a}$ and $\eta\in
F_{q^{2}}$ be a primitive $r$th root of unity. In this section, we
consider $\eta$-constacyclic codes over $F_{q^{2}}$ of length $n$ to
construct quantum codes. To do this, we give a sufficient condition
for $\eta$-constacyclic codes over $F_{q^{2}}$ of length $n$ which
contain their Hermitian duals. First, we compute $q^{2}$-cyclotomic
cosets modulo $rn$.

 Let $r=q+1$. we compute $q^{2}$-cyclotomic cosets modulo $(q+1)n$.

\noindent{\bf  Lemma 3.1:}  Let $q=2am+t$, $a|t^{2}+1$ and
$n=\frac{q^{2}+1}{a}$, where $a$ is a odd. Let
$s=\frac{q^{2}+1}{2}$. Then, for any integer $i\in
\Omega=\{1+(q+1)j|0\leq j\leq n-1\}$, the $q^{2}$-cyclotomic coset
$C_{i}$ modulo $(q+1)n$ is given by

1) $C_{s}=\{s\}$ and $C_{s+n(q+1)/2}=\{s+n(q+1)/2\}$.

2) $C_{s-(q+1)j}=\{s-(q+1)j,s+(q+1)j\}$ for $1\leq j \leq
\frac{n}{2}-1$.

\noindent{\it Proof.} (1) If $j=(q-1)/2$, then $1+(q+1)j=s$. This
means that $s$ must be in $\Omega$. Moreover,
$[s+n(q+1)/2]q^{2}=sq^{2}+n(q+1)(q^{2}-1)/2+n(q+1)/2\equiv
s+n(q+1)/2$ mod $(q+1)n$. Hence, $C_{s}=\{s\}$ and
$C_{s+n(q+1)/2}=\{s+n(q+1)/2\}$.

(2) the proof is the similarly as Lemma 3.12 in Ref.(Kai2014).

Then, let $r=1$, we compute $q^{2}$-cyclotomic cosets modulo $n$.

\noindent{\bf  Lemma 3.2:}  Let $n=\frac{q^{2}+1}{a}$ and $a$ is an
odd. Then, for any integer $i\in \Omega=\{1+j|0\leq j\leq n-1\}$,
the $q^{2}$-cyclotomic coset $C_{i}$ modulo $n$ is given by
$C_{i}=\{i,n-i\}$ for $1\leq i \leq \frac{n}{2}-1$.

\noindent{\it Proof.}  $iq^{2}=\frac{ai(q^{2}+1)}{a}-i$.
$\frac{ai(q^{2}+1)}{a}-i=\frac{(ai-1)(q^{2}+1)}{a}+\frac{(q^{2}+1)}{a}-i$
$\equiv n-i$ mod $n$. Moreover,  $(n-i)q^{2}\equiv i$ mod $n$.
Hence,  $C_{i}=\{i,n-i\}$ for $1\leq i \leq \frac{n}{2}-1$.

\subsection{ New  Quantum MDS Codes of Lenght
$n=\frac{q^{2}+1}{13}$}

Let $a=13$, $q$ be an odd prime power with the form $26m+5$ or
$26m+21$, where $m$ is a positive integer. Then $13$ is a divisor of
$q^{2}+1$. Let $n=\frac{q^{2}+1}{13}$, $r=q+1$ and $\eta\in
F_{q^{2}}$ be a primitive $r$th root of unity.  First, we consider
$\eta$-constacyclic codes over $F_{q^{2}}$ of length $n$ to
construct quantum codes. To do this, we give a sufficient condition
for $\eta$-constacyclic codes over $F_{q^{2}}$ of length $n$ which
contain their Hermitian duals.

\noindent{\bf  Lemma 3.3:} Let $q$ be an odd prime power,
$n=\frac{q^{2}+1}{13}$ and $s=\frac{q^{2}+1}{2}$. Suppose
$\mathcal{C}$ is a $q^{2}$-ary constacyclic code of length $n$  with
define  set $T=\bigcup_{i=0}^{\delta}C_{s+(q+1)i}$. For $m$ is a
positive integer, then

  (1) if $q$ is the form $26m+5$, and  $0\leq \delta\leq 5m$, then $\mathcal{C}$$^{\perp_{h}}$ $\subseteq \mathcal{C}$.

(2)  if $q$ is the form $26m+21$,  and  $0\leq \delta\leq 5m+3$, then $\mathcal{C}$$^{\perp_{h}}$ $\subseteq \mathcal{C}$.

\noindent{\it Proof.} According to Lemma 2.2 and Lemma 2.3, one obtain
that $\mathcal{C}$$^{\perp_{h}}$ $\subseteq \mathcal{C}$ if and only if there is no  skew symmetric cyclotomic  coset and any two  cyclotomic  cosets
do not form a skew asymmetric pair in the defining set $T$.

(1) For $q$ is the form $26m+5$, $n=\frac{q^{2}+1}{13}=52m^{2}+20m+2$,
$s=\frac{q^{2}+1}{2}=338m^{2}+130m+13$.
Let $x,y \in \Omega=\{s+(q+1)j|0\leq j\leq 5m\}$. one only testify that $x+yq \not\equiv 0$ mod $rn$ holds.
To do this, we divide $\Omega$ into three parts such that $\Omega=\bigcup_{i=1}^{3} I_{i}$, where
$I_{1}=[s,s+(q+1)m]$, $I_{2}=[s+(q+1)(m+1),s+(q+1)3m]$ and $I_{3}=[s+(q+1)(3m+1),s+(q+1)5m]$.

First, we testify that any cyclotomic  coset is not skew symmetric.
If $x\in I_{1}$, $6rn<\frac{(q^{2}+1)(q+1)}{26}+6rn=s(q+1)\leq x(q+1)$
$\leq [s+(q+1)m](q+1)= \frac{(q^{2}+1)(q+1)}{2}+(q^{2}+1)m$.
Since, $rn=(q+1)(52m^{2}+20m+2)=1352m^{3}+832m^{2}+172m+12$,
$s(q+1)=6rn+\frac{(q^{2}+1)(q+1)}{26}$
and $(q^{2}+1)m=676m^{3}+312m^{2}+36m$, then $6rn<x(q+1)<7rn$.
If $x\in I_{2}$, $7rn<6rn+r(52m^{2}+41m+6)=\frac{(q^{2}+1)(q+1)}{26}+6rn+(q+1)^{2}(m+1)=[s+(q+1)(m+1)](q+1)\leq x(q+1)$$\leq [s+(q+1)(3m)](q+1)=6rn+r(104m^{2}+25m+1)<8rn$.
If $x\in I_{3}$, $8rn<6rn+r(104m^{2}+51m+6)=[s+(q+1)(3m+1)](q+1)\leq x(q+1)$$\leq [s+(q+1)(5m)](q+1)=6rn+r(156m^{2}+35m+1)<9rn$.
Hence, there is no
skew symmetric cyclotomic  coset in the defining set $T$.

Then, we testify that any two  cyclotomic  cosets
do not form a skew asymmetric pair.
If $x,y\in I_{1}$, $6rn<\frac{(q^{2}+1)(q+1)}{26}+6rn\leq x+yq$
$\leq [s+(q+1)m](q+1)<7rn$; If $x,y\in I_{2}$, $7rn<6rn+r(52m^{2}+41m+6)\leq x+yq$$\leq [s+(q+1)(3m)](q+1)=6rn+r(104m^{2}+25m+1)<8rn$.
If $x,y\in I_{3}$, $8rn<6rn+r(104m^{2}+51m+6)\leq x+yq$$\leq [s+(q+1)(5m)](q+1)=6rn+r(156m^{2}+35m+1)<9rn$.
And If $x\in I_{1}\cup I_{2}$, $y\in I_{1}$,
$6rn<\frac{(q^{2}+1)(q+1)}{26}+6rn\leq x+yq$
$\leq s+(q+1)3m+[s+(q+1)m]q=(q+1)[s+m(q+3)]=r(364m^{2}+138m+13)<7rn$;
and If $x\in I_{1}\cup I_{2}\cup I_{3}$, $y\in I_{1}$,
$6rn<\frac{(q^{2}+1)(q+1)}{26}+6rn\leq x+yq$
$\leq s+(q+1)5m+[s+(q+1)m]q=(q+1)[s+m(q+3)]=r(364m^{2}+140m+13)<7rn$;
and If $x\in I_{2}\cup I_{3}$, $y\in I_{2}$,
$7rn<6rn+r(52m^{2}+41m+6)\leq x+yq$$\leq  s+(q+1)5m+[s+(q+1)3m]q=(q+1)[416m^{2}+150m+13]<8rn$.

Hence, any two  cyclotomic  cosets
do not form a skew asymmetric pair. in the defining set $T$.
Therefore, if $q$ is the form $26m+5$, and  $0\leq \delta\leq 5m$, then $\mathcal{C}$$^{\perp_{h}}$ $\subseteq \mathcal{C}$.

(2) For $q$ is the form $26m+21$.  $n=\frac{q^{2}+1}{13}=52m^{2}+84m+34$,
$s=\frac{q^{2}+1}{2}=338m^{2}+546m+221$.
  Let $x,y \in \Omega=\{s+(q+1)j|0\leq j\leq 5m+3\}$. one only testify that $x+yq \not\equiv 0$ mod $rn$ holds.
To do this, we divide $\Omega$ into three parts such that $\Omega=\bigcup_{i=1}^{3} I_{i}$, where
$I_{1}=[s,s+(q+1)m]$, $I_{2}=[s+(q+1)(m+1),s+(q+1)(3m+2)]$ and $I_{3}=[s+(q+1)(3m+3),s+(q+1)(5m+3)]$.

First, we testify that any cyclotomic  coset is not skew symmetric.
If $x\in I_{1}$, $6rn<\frac{(q^{2}+1)(q+1)}{26}+6rn=s(q+1)\leq x(q+1)$
$\leq [s+(q+1)m](q+1)= 6rn+52m^{2}+64m+17<7rn$.
If $x\in I_{2}$, $7rn<6rn+r(52m^{2}+90m+39)=\frac{(q^{2}+1)(q+1)}{26}+6rn+(q+1)^{2}(m+1)=[s+(q+1)(m+1)](q+1)\leq x(q+1)$$\leq [s+(q+1)(3m+3)](q+1)=6rn+r(104m^{2}+160m+61)<8rn$.
If $x\in I_{3}$, $8rn<8rn+r(18m+15)=[s+(q+1)(3m+3)](q+1)\leq x(q+1)$$\leq [s+(q+1)(5m+3)](q+1)=8rn+r(52m^{2}+18m+15)<9rn$.
Hence, there is no
skew symmetric cyclotomic  coset in the defining set $T$.

Then, we testify that any two  cyclotomic  cosets
do not form a skew asymmetric pair.

If $x,y\in I_{1}$, $6rn<\frac{(q^{2}+1)(q+1)}{26}+6rn\leq x+yq$
$\leq [s+(q+1)m](q+1)<7rn$; If $x,y\in I_{2}$, $7rn<6rn+r(52m^{2}+90m+39)\leq x+yq$$\leq [s+(q+1)(3m+2)](q+1)=6rn+r(104m^{2}+160m+61)<8rn$.
If $x,y\in I_{3}$, $8rn<8rn+r(18m+15)\leq x+yq$$\leq [s+(q+1)(5m+3)](q+1)=8rn+r(52m^{2}+18m+15)<9rn$.
And If $x\in I_{1}\cup I_{2}$, $y\in I_{1}$,
$6rn<\frac{(q^{2}+1)(q+1)}{26}+6rn\leq x+yq$
$\leq s+(q+1)(3m+2)+[s+(q+1)m]q=(q+1)[s+m(q+3)+2]=6rn+r(52m^{2}+56m+19)<7rn$;
and If $x\in I_{1}\cup I_{2}\cup I_{3}$, $y\in I_{1}$,
$6rn<\frac{(q^{2}+1)(q+1)}{26}+6rn\leq x+yq$
$\leq s+(q+1)(5m+3)+[s+(q+1)m]q=(q+1)[s+m(q+3)]=6rn+r(52m^{2}+68m+24)<7rn$;
and If $x\in I_{2}\cup I_{3}$, $y\in I_{2}$,
$7rn<6rn+r(52m^{2}+90m+39)\leq x+yq$$\leq  s+(q+1)(5m+3)+[s+(q+1)(3m+2)]q=7rn+r(52m^{2}+78m+27)<8rn$.

Hence, any two  cyclotomic  cosets do not form a skew asymmetric
pair in the defining set $T$. Therefore, if $q$ is the form
$26m+21$, and  $0\leq \delta\leq 5m+3$, then
$\mathcal{C}$$^{\perp_{h}}$ $\subseteq \mathcal{C}$.

\noindent{\bf  Theory 3.4:} If $q$ is an odd prime power of the form
$26m+5$ or $26m+21$,
 then there exists a q-ary $[[\frac{q^{2}+1}{13},\frac{q^{2}+1}{13}-2d+2,d]]$-quantum MDS
codes, where $2\leq d \leq 10m+2$ is even for $q$ with the form of $26m+5$;
 and $2\leq d \leq 10m+8$ is even for $q$ with the form of $26m+21$.

\noindent{\it Proof.} Fix $s=\frac{q^{2}+1}{2}$. Consider the
$\eta$-constacyclic codes over $F_{q^{2}}$ of length
$n=\frac{q^{2}+1}{13}$ with defining set
$T=\bigcup_{i=0}^{\delta}C_{s+(q+1)i}$, where $0\leq \delta\leq 5m$
for $q$ with the form of $26m+5$, and $0\leq \delta\leq 5m+3$ for
$q$ with the form of $26m+21$. By Lemma 3.3, there is
$\mathcal{C}$$^{\perp_{h}}$ $\subseteq \mathcal{C}$. Form lemma 3.1,
we can obtain that $T$ consists of $2\delta+1$ integers
$\{s-(q+1)\delta,\cdots,s-(q+1),s,s+(q+1),\cdots,s+(q+1)\delta\}$.
It implies that $\mathcal{C}$ has minimum distance at least
$2\delta+2$ and the minimum distance $d_{C}$ is even. Hence,
$\mathcal{C}$ is a $q^{2}$-ary $\eta$ constacyclic code with
parameters $[n,n-(2\delta-1),\geq 2\delta+2]$. Combining the
Hermitian construction with quantum Singleton bound, we can obtain a
quantum MDS code with parameters $[[n,n-4\delta-2,2\delta+2]]_{q}$.
Hence, we can obtian $q$
ary-$[[\frac{q^{2}+1}{13},\frac{q^{2}+1}{13}-2d+2,d]]$ quantum MDS
code, where $2\leq d \leq 10m+2$ is even for $q$ with the form of
$26m+5$ and $2\leq d \leq 10m+8$ is even for  $q$ is the form of
$26m+21$.

\begin{center}
Table 1 Quantum MDS codes with length $n=\frac{q^{2}+1}{13}$\\
\begin{tabular}{llllllll}
  \hline
  % after \\: \hline or \cline{col1-col2} \cline{col3-col4} ...
  Quantum MDS code       & the form of q      & m     & q                 &r    &&d  \\
\hline
  $[[74,76-2d,d]]_{31}$   & 26m+5             & 1       &$31$            &32  &&$2\leq d\leq 12$ is even \\

 $[[530,532-2d,d]]_{83}$   & 26m+5             & 3       &$83$            &84  &&$2\leq d\leq 32$ is even \\

 $[[170,172-2d,d]]_{47}$   &26m+21            &1        &$47$          &48  &&$2\leq d\leq 18$ is even \\

 $[[410,412-2d,d]]_{73}$   &26m+21            &2        &$73$          &74  &&$2\leq d\leq 28$ is even \\

  \hline
  \end{tabular}
\end{center}

\subsection{ New  Quantum MDS Codes of Lenght
$n=\frac{q^{2}+1}{17}$}

Let $a=17$, $q$ be an odd prime power with the form $34m+13$ or
$34m+21$, where $m$ is a positive integer. Then $17$ is a divisor of
$q^{2}+1$. Let $n=\frac{q^{2}+1}{17}$, $r=q+1$ and $\eta\in
F_{q^{2}}$ be a primitive $r$th root of unity.  First, we consider
$\eta$-constacyclic codes over $F_{q^{2}}$ of length $n$ to
construct quantum codes. To do this, we give a sufficient condition
for $\eta$-constacyclic codes over $F_{q^{2}}$ of length $n$ which
contain their Hermitian duals.

\noindent{\bf  Lemma 3.5:} Let $q$ be an odd prime power,
$n=\frac{q^{2}+1}{17}$ and $s=\frac{q^{2}+1}{2}$. Suppose
$\mathcal{C}$ is a $q^{2}$-ary constacyclic code of length $n$  with
define  set $T=\bigcup_{i=0}^{\delta}C_{s+(q+1)i}$. For $m$ is a
positive integer, then

  (1) if $q$ is the form $34m+13$, and  $0\leq \delta\leq 5m+1$, then $\mathcal{C}$$^{\perp_{h}}$ $\subseteq \mathcal{C}$.

(2)  if $q$ is the form $34m+21$,  and  $0\leq \delta\leq 5m+2$, then $\mathcal{C}$$^{\perp_{h}}$ $\subseteq \mathcal{C}$.

\noindent{\it Proof.} According to Lemma 2.2 and Lemma 2.3, one
obtain that $\mathcal{C}$$^{\perp_{h}}$ $\subseteq \mathcal{C}$ if
and only if there is no  skew symmetric cyclotomic  coset and any
two  cyclotomic  cosets do not form a skew asymmetric pair in the
defining set $T$.  For $q$ is the form $34m+13$,
$n=\frac{q^{2}+1}{17}=68m^{2}+52m+10$,
$s=\frac{q^{2}+1}{2}=8n+\frac{n}{2}$. Let $x,y \in
\Omega=\{s+(q+1)j|0\leq j\leq 5m\}$. one only testify that $x+yq
\not\equiv 0$ mod $rn$ holds. To do this, we divide $\Omega$ into
three parts such that $\Omega=\bigcup_{i=1}^{3} I_{i}$, where
$I_{1}=[s,s+(q+1)m]$, $I_{2}=[s+(q+1)(m+1),s+(q+1)(3m+1)]$ and
$I_{3}=[s+(q+1)(3m+2),s+(q+1)(5m+1)]$.

First, we testify that any cyclotomic  coset is not skew symmetric.
If $x\in I_{1}$, $8rn<\frac{(q^{2}+1)(q+1)}{34}+8rn=s(q+1)\leq
x(q+1)$ $\leq [s+(q+1)m](q+1)=8rn+\frac{(q^{2}+1)(q+1)}{34}<9rn$; If
$x\in I_{2}$, $9rn<8rn+(68m^{2}+74m+19)r=(s+(q+1)m)(q+1)\leq x(q+1)$
$\leq [s+(q+1)(3m+1)](q+1)\leq 8rn+(136m^{2}+102m+19)r<10rn$;
 If $x\in I_{3}$,
$10rn<8rn+136m^{2}+136m+33=(s+(q+1)(3m+2))(q+1)\leq x(q+1)$ $\leq
[s+(q+1)(5m+1)](q+1)<11rn$. Hence, there is no skew symmetric
cyclotomic  coset in the defining set $T$. Using the same method,
one can testy that any two  cyclotomic  cosets do not form a skew
asymmetric pair. Therefore, if $q$ is the form $34m+13$, and $0\leq
\delta\leq 5m+1$, then $\mathcal{C}$$^{\perp_{h}}$ $\subseteq
\mathcal{C}$. And for $q$ is the form $34m+21$, one also can testy
that there is no skew asymmetric pairs in defining set $T$ and
$\mathcal{C}$$^{\perp_{h}}$ $\subseteq \mathcal{C}$.

\noindent{\bf  Theory 3.6:} If $q$ is an odd prime power of the form
$34m+13$ or $34m+21$,
 then there exists a q-ary $[[\frac{q^{2}+1}{17},\frac{q^{2}+1}{17}-2d+2,d]]$-quantum MDS
codes, where $2\leq d \leq 10m+4$ is even for $q$ with the form of $34m+13$;
 and $2\leq d \leq 10m+6$ is even for $q$ with the form of $34m+21$.

\noindent{\it Proof.} Fix $s=\frac{q^{2}+1}{2}$. Consider the
$\eta$-constacyclic codes over $F_{q^{2}}$ of length
$n=\frac{q^{2}+1}{17}$ with defining set
$T=\bigcup_{i=0}^{\delta}C_{s+(q+1)i}$, where $0\leq \delta\leq
5m+1$ for $q$ with the form of $34m+13$, and $0\leq \delta\leq 5m+2$
for $q$ with the form of $34m+21$. By Lemma 3.5, there is
$\mathcal{C}$$^{\perp_{h}}$ $\subseteq \mathcal{C}$. Form lemma 3.1,
we can obtain that $T$ consists of $2\delta+1$ integers
$\{s-(q+1)\delta,\cdots,s-(q+1),s,s+(q+1),\cdots,s+(q+1)\delta\}$.
It implies that $\mathcal{C}$ has minimum distance at least
$2\delta+2$ and the minimum distance $d_{C}$ is even. Hence,
$\mathcal{C}$ is a $q^{2}$-ary $\eta$ constacyclic code with
parameters $[n,n-(2\delta-1),\geq 2\delta+2]$.
 Combining the Hermitian construction with quantum Singleton
bound, we can obtain a quantum MDS code with parameters $[[n,n-4\delta-2,2\delta+2]]_{q}$.
Hence, we can obtian $q$ ary-$[[\frac{q^{2}+1}{17},\frac{q^{2}+1}{17}-2d+2,d]]$ quantum MDS code, where
$2\leq d \leq 10m+4$ is even for $q$ with the form of $34m+13$
and $2\leq d \leq 10m+6$ is even for  $q$ is the form of $34m+21$.

\begin{center}
Table 2 Quantum MDS codes with length $n=\frac{q^{2}+1}{17}$\\
\begin{tabular}{llllllllll}
  \hline
  % after \\: \hline or \cline{col1-col2} \cline{col3-col4} ...
  Quantum MDS code            & the form of q       &m  &q       &r     & &d  \\
\hline
  $[[130,132-2d,d]]_{47}$     &34m+13               &1  &47     &48   &&$2\leq d\leq 14$ is even \\

 $[[466,468-2d,d]]_{89}$      &34m+21               &2   &89      &90   &&$2\leq d\leq 26$ is even \\

  \hline
  \end{tabular}
\end{center}

\subsection{ New  Quantum MDS Codes of Lenght
$n=\frac{q^{2}+1}{25}$}

Let $a=25$, $q$ be an odd prime power with the form $50m+7$ or
$50m+43$, where $m$ is a positive integer. Then $25$ is a divisor of
$q^{2}+1$. Let $n=\frac{q^{2}+1}{25}$, $r=q+1$ and $\eta\in
F_{q^{2}}$ be a primitive $r$th root of unity.  We consider
$\eta$-constacyclic codes over $F_{q^{2}}$ of length $n$ to
construct quantum codes. First, we give a sufficient condition for
$\eta$-constacyclic codes over $F_{q^{2}}$ of length $n$ which
contain their Hermitian duals.

\noindent{\bf  Lemma 3.7:} Let $q$ be an odd prime power,
$n=\frac{q^{2}+1}{25}>q$. Suppose $\mathcal{C}$ is a $q^{2}$-ary
constacyclic code of length $n$  with define  set
$T=\bigcup_{i=0}^{\delta}C_{s+(q+1)i}$. For $m\geq 0$ is a positive
integer, then

(1)  if $q$ is the form $50m+7$,  and  $0\leq \delta\leq 7m$, then $\mathcal{C}$$^{\perp_{h}}$ $\subseteq \mathcal{C}$.

(2) if $q$ is the form $50m+43$, and  $0\leq \delta\leq 7m+5$, then $\mathcal{C}$$^{\perp_{h}}$ $\subseteq \mathcal{C}$.

\noindent{\it Proof.} According to Lemma 2.2 and Lemma 2.3, one obtain
that $\mathcal{C}$$^{\perp_{h}}$ $\subseteq \mathcal{C}$ if and only if there is no
skew symmetric cyclotomic  coset and any two  cyclotomic  cosets
do not form a skew asymmetric pair in the defining set $T$.

 For $q$ is the form $50m+7$,
$n=\frac{q^{2}+1}{25}=100m^{2}+28m+2$,
$s=\frac{q^{2}+1}{2}=12n+\frac{n}{2}$. Let $x,y \in
\Omega=\{s+(q+1)j|0\leq j\leq 7m\}$. one only testify that $x+yq
\not\equiv 0$ mod $rn$ holds. To do this, we divide $\Omega$ into
three parts such that $\Omega=\bigcup_{i=1}^{4} I_{i}$, where
$I_{1}=[s,s+(q+1)m]$, $I_{2}=[s+(q+1)(m+1),s+(q+1)3m]$,
$I_{3}=[s+(q+1)(3m+1),s+(q+1)(5m+1)]$ and
$I_{3}=[s+(q+1)(5m+2),s+(q+1)7m]$.

First, we testify that any cyclotomic  coset is not skew symmetric.
If $x\in I_{1}$, $12rn<\frac{(q^{2}+1)(q+1)}{50}+12rn=s(q+1)\leq
x(q+1)$ $\leq [s+(q+1)m](q+1)= 12rn+(100m^{2}+22m+1)r<13rn$; If
$x\in I_{2}$, $13rn<12rn+(100m^{2}+58m+8)r=(s+(q+1)(m+1))(q+1)\leq
x(q+1)$ $\leq [s+(q+1)3m](q+1)\leq 12rn+(200m^{2}+38m+1)r<14rn$;
 If $x\in I_{3}$,
$14rn<12rn+(200m^{2}+88m+9)r=(s+(q+1)(3m+1))(q+1)\leq x(q+1)$ $\leq
[s+(q+1)(5m+1)](q+1)<15rn$;  If $x\in I_{4}$,
$15rn<(s+(q+1)(3m+1))(q+1)\leq x(q+1)$ $\leq
[s+(q+1)7m](q+1)<16rn$;. Hence, there is no skew symmetric
cyclotomic  coset in the defining set $T$. Using the same method,
one can testy that any two cyclotomic  cosets do not form a skew
asymmetric pair. Therefore, if $q$ is the form $50m+7$, and $0\leq
\delta\leq 7m$, then $\mathcal{C}$$^{\perp_{h}}$ $\subseteq
\mathcal{C}$. And for $q$ is the form $50m+41$, using the same
method, one also can testy that there is skew symmetric cyclotomic
coset and any two  cyclotomic  cosets do not form a skew asymmetric
pair in the defining set $T$. Therefore, $\mathcal{C}$$^{\perp_{h}}$
$\subseteq \mathcal{C}$.

\noindent{\bf  Theory 3.8:} If $q$ is an odd prime power of the form
$50m+7$ or $50m+43$,
 then there exists a q-ary $[[\frac{q^{2}+1}{25},\frac{q^{2}+1}{25}-2d+2,d]]$-quantum MDS
codes, where $2\leq d \leq 14m+2$ is even for $q$ with the form of $50m+7$;
 and $2\leq d \leq 14m+12$ is even for $q$ with the form of $50m+43$.

\noindent{\it Proof.} Fix $s=\frac{q^{2}+1}{2}$. Consider the
$\eta$-constacyclic codes over $F_{q^{2}}$ of length
$n=\frac{q^{2}+1}{25}$ with defining set
$T=\bigcup_{i=0}^{\delta}C_{s+(q+1)i}$, where $0\leq \delta\leq 7m$
for $q$ with the form of $50m+7$, and $0\leq \delta\leq 7m+5$ for
$q$ with the form of $50m+43$. By Lemma 3.7, there is
$\mathcal{C}$$^{\perp_{h}}$ $\subseteq \mathcal{C}$. Form lemma 3.1,
we can obtain that $T$ consists of $2\delta+1$ integers
$\{s-(q+1)\delta,\cdots,s-(q+1),s,s+(q+1),\cdots,s+(q+1)\delta\}$.
It implies that $\mathcal{C}$ has minimum distance at least
$2\delta+2$ and the minimum distance $d_{C}$ is even. Hence,
$\mathcal{C}$ is a $q^{2}$-ary $\eta$ constacyclic code with
parameters $[n,n-(2\delta-1),\geq 2\delta+2]$. Combining the
Hermitian construction with quantum Singleton bound, we can obtain a
quantum MDS code with parameters $[[n,n-4\delta-2,2\delta+2]]_{q}$.
Hence, we can obtian $q$
ary-$[[\frac{q^{2}+1}{25},\frac{q^{2}+1}{25}-2d+2,d]]$ quantum MDS
code,
 where $2\leq d \leq 14m+10$ is even for $q$ with the form of $50m+43$ and $2\leq d \leq 14m+2$ is even for
  $q$ is the form of $50m+7$.

\begin{center}
Table 3 Quantum MDS codes with length $n=\frac{q^{2}+1}{25}$\\
\begin{tabular}{llllllllllll}
  \hline
  % after \\: \hline or \cline{col1-col2} \cline{col3-col4} ...
  Quantum MDS code            &the form of q        &m   &q   &r     &&d  \\
\hline
  $[[458,460-2d,d]]_{107}$     &50m+7               &2    &107    &108   &&$2\leq d\leq 30$ is even \\

 $[[74,76-2d,d]]_{43}$        &50m+43               &0    &43      &44   &&$2\leq d\leq 12$  is even \\

  \hline
  \end{tabular}
\end{center}

\subsection{ New  Quantum MDS Codes of Lenght
$n=\frac{q^{2}+1}{29}$}

Let $a=29$ and $q$ be an odd prime power with the form $58m+17$ or
$58m+41$, where $m$ is a positive integer. Then $29$ is a divisor of
$q^{2}+1$. Let $n=\frac{q^{2}+1}{29}$, $r=q+1$ and $\eta\in
F_{q^{2}}$ be a primitive $r$th root of unity.  We consider
$\eta$-constacyclic codes over $F_{q^{2}}$ of length $n$ to
construct quantum codes. First, we give a sufficient condition for
$\eta$-constacyclic codes over $F_{q^{2}}$ of length $n$ which
contain their Hermitian duals.

\noindent{\bf  Lemma 3.9:} Let $q$ be an odd prime power,
$n=\frac{q^{2}+1}{29}>q$. Suppose $\mathcal{C}$ is a $q^{2}$-ary
constacyclic code of length $n$  with define  set
$T=\bigcup_{i=0}^{\delta}C_{s+(q+1)i}$. For $m\geq 0$ is a positive
integer, then

(1)  if $q$ is the form $58m+17$,  and  $0\leq \delta\leq 7m+1$, then $\mathcal{C}$$^{\perp_{h}}$ $\subseteq \mathcal{C}$.

(2) if $q$ is the form $58m+41$, and  $0\leq \delta\leq 7m+4$, then $\mathcal{C}$$^{\perp_{h}}$ $\subseteq \mathcal{C}$.

\noindent{\it Proof.} According to Lemma 2.2 and Lemma 2.3, one obtain
that $\mathcal{C}$$^{\perp_{h}}$ $\subseteq \mathcal{C}$ if and only if there is no  skew symmetric cyclotomic  coset and any two  cyclotomic  cosets
do not form a skew asymmetric pair in the defining set $T$.
 For $q$ is the form $50m+7$,
$n=\frac{q^{2}+1}{29}=116m^{2}+68m+10$,
$s=\frac{q^{2}+1}{2}=14n+\frac{n}{2}$. Let $x,y \in
\Omega=\{s+(q+1)j|0\leq j\leq 7m+1\}$. one only testify that $x+yq
\not\equiv 0$ mod $rn$ holds. To do this, we divide $\Omega$ into
four parts such that $\Omega=\bigcup_{i=1}^{4} I_{i}$, where
$I_{1}=[s,s+(q+1)m]$, $I_{2}=[s+(q+1)(m+1),s+(q+1)(3m+1)]$,
$I_{3}=[s+(q+1)(3m+2),s+(q+1)(5m+1)]$ and
$I_{4}=[s+(q+1)(5m+2),s+(q+1)(7m+1)]$. First, we testify that any
cyclotomic  coset is not skew symmetric. If $x\in I_{1}$,
$14rn<\frac{(q^{2}+1)(q+1)}{58}+14rn=s(q+1)\leq x(q+1)$ $\leq
[s+(q+1)m](q+1)= 14rn+(116m^{2}+52m+5)r<15rn$; If $x\in I_{2}$,
$15rn<14rn+(116m^{2}+76m+18)r=(s+(q+1)(m+1))(q+1)\leq x(q+1)$ $\leq
[s+(q+1)(3m+1)](q+1)\leq 14rn+(232m^{2}+146m+23)r<16rn$;
 If $x\in I_{3}$,
$16rn<16rn+(68m+21)r=(s+(q+1)(3m+1))(q+1)\leq x(q+1)$ $\leq
[s+(q+1)(5m+1)](q+1)=17rn-(22m+7)<17rn$;  If $x\in I_{4}$,
$17rn<17rn+(36m+11)r=(s+(q+1)(5m+2))(q+1)\leq x(q+1)$ $\leq
[s+(q+1)(7m+1)](q+1)=18rn-(54m+23)r<18rn$;. Hence, there is no skew
symmetric cyclotomic  coset in the defining set $T$. Using the same
method, one can testy that any two cyclotomic  cosets do not form a
skew asymmetric pair. Therefore, if $q$ is the form $58m+17$, and
$0\leq \delta\leq 7m+1$, then $\mathcal{C}$$^{\perp_{h}}$ $\subseteq
\mathcal{C}$. And for $q$ is the form $58m+41$, using the same
method, one also can testy that there is skew symmetric cyclotomic
coset and any two  cyclotomic cosets do not form a skew asymmetric
pair in the defining set $T$. Therefore, $\mathcal{C}$$^{\perp_{h}}$
$\subseteq \mathcal{C}$.

\noindent{\bf  Theory 3.10:} If $q$ is an odd prime power of the
form $58m+17$ or $58m+41$,
 then there exists a q-ary $[[\frac{q^{2}+1}{29},\frac{q^{2}+1}{29}-2d+2,d]]$-quantum MDS
codes, where $2\leq d \leq 14m+4$ is even for $q$ with the form of $58m+17$;
 and $2\leq d \leq 14m+10$ is even for $q$ with the form of $58m+41$.

\noindent{\it Proof.} Fix $s=\frac{q^{2}+1}{2}$. Consider the
$\eta$-constacyclic codes over $F_{q^{2}}$ of length
$n=\frac{q^{2}+1}{29}$ with defining set
$T=\bigcup_{i=0}^{\delta}C_{s+(q+1)i}$, where $0\leq \delta\leq
7m+2$ for $q$ with the form of $58m+17$, and $0\leq \delta\leq 7m+4$
for $q$ with the form of $58m+41$. By Lemma 3.9, there is
$\mathcal{C}$$^{\perp_{h}}$ $\subseteq \mathcal{C}$. Form lemma 3.1,
we can obtain that $T$ consists of $2\delta+1$ integers
$\{s-(q+1)\delta,\cdots,s-(q+1),s,s+(q+1),\cdots,s+(q+1)\delta\}$.
It implies that $\mathcal{C}$ has minimum distance at least
$2\delta+2$ and the minimum distance $d_{C}$ is even. Hence,
$\mathcal{C}$ is a $q^{2}$-ary $\eta$ constacyclic code with
parameters $[n,n-(2\delta-1),\geq 2\delta+2]$. Combining the
Hermitian construction with quantum Singleton bound, we can obtain a
quantum MDS code with parameters $[[n,n-4\delta-2,2\delta+2]]_{q}$.
Hence, we can obtian $q$
ary-$[[\frac{q^{2}+1}{25},\frac{q^{2}+1}{29}-2d+2,d]]$ quantum MDS
code, where $2\leq d \leq 14m+4$ is even for $q$ with the form of
$58m+17$ and $2\leq d \leq 14m+10$ is even for  $q$ is the form of
$58m+41$.

\begin{center}
Table 4 Quantum MDS codes with length $n=\frac{q^{2}+1}{29}$\\
\begin{tabular}{llllllllll}
  \hline
  % after \\: \hline or \cline{col1-col2} \cline{col3-col4} ...
  Quantum MDS code            &the form of q     &m &q   &r     &&d  \\
\hline
  $[[1258,1260-2d,d]]_{191}$  &58m+17            &2  &191     &192   &&$2\leq d\leq 46$ is even \\

 $[[58,60-2d,d]]_{41}$        &58m+41           &0  &41    &42   &&$2\leq d\leq 10$   is even \\

  \hline
  \end{tabular}
\end{center}

\subsection{ New  Quantum MDS Codes of Lenght
$n=\frac{q^{2}+1}{37}$}

Let $a=37$ and $q$ be an odd prime power with the form $74m+31$ or
$74m+43$, where $m$ is a positive integer. Then $37$ is a divisor of
$q^{2}+1$. Let $n=\frac{q^{2}+1}{37}$, $r=q+1$ and $\eta\in
F_{q^{2}}$ be a primitive $r$th root of unity.  We consider
$\eta$-constacyclic codes over $F_{q^{2}}$ of length $n$ to
construct quantum codes. First, we give a sufficient condition for
$\eta$-constacyclic codes over $F_{q^{2}}$ of length $n$ which
contain their Hermitian duals.

\noindent{\bf  Lemma 3.11:} Let $q$ be an odd prime power,
$n=\frac{q^{2}+1}{37}>q$. Suppose $\mathcal{C}$ is a $q^{2}$-ary
constacyclic code of length $n$  with define  set
$T=\bigcup_{i=0}^{\delta}C_{s+(q+1)i}$. For $m\geq 0$ is a positive
integer, then

(1)  if $q$ is the form $74m+31$,  and  $0\leq \delta\leq 7m+2$,
then $\mathcal{C}$$^{\perp_{h}}$ $\subseteq \mathcal{C}$.

(2) if $q$ is the form $74m+43$, and  $0\leq \delta\leq 7m+3$, then $\mathcal{C}$$^{\perp_{h}}$ $\subseteq \mathcal{C}$.

\noindent{\it Proof.} According to Lemma 2.2 and Lemma 2.3, one obtain
that $\mathcal{C}$$^{\perp_{h}}$ $\subseteq \mathcal{C}$ if and only if there is no  skew symmetric cyclotomic  coset and any two  cyclotomic  cosets
do not form a skew asymmetric pair in the defining set $T$.
 For $q$ is the form $74m+31$,
$n=\frac{q^{2}+1}{37}=148m^{2}+124m+26$,
$s=\frac{q^{2}+1}{2}=18n+\frac{n}{2}$. Let $x,y \in
\Omega=\{s+(q+1)j|0\leq j\leq 7m+2\}$. one only testify that $x+yq
\not\equiv 0$ mod $rn$ holds. To do this, we divide $\Omega$ into
four parts such that $\Omega=\bigcup_{i=1}^{4} I_{i}$, where
$I_{1}=[s,s+(q+1)m]$, $I_{2}=[s+(q+1)(m+1),s+(q+1)(3m+1)]$,
$I_{3}=[s+(q+1)(3m+2),s+(q+1)(5m+1)]$ and
$I_{4}=[s+(q+1)(5m+2),s+(q+1)(7m+1)]$. First, we testify that any
cyclotomic  coset is not skew symmetric. If $x\in I_{1}$,
$18rn<\frac{(q^{2}+1)(q+1)}{74}+18rn=s(q+1)\leq x(q+1)$ $\leq
[s+(q+1)m](q+1)= 18rn+(148m^{2}+94m+13)r<19rn$; If $x\in I_{2}$,
$19rn<19rn+(44m+19)r=(s+(q+1)(m+1))(q+1)\leq x(q+1)$ $\leq
[s+(q+1)(3m+1)](q+1)\leq 20rn-(18m-6)r<20rn$;
 If $x\in I_{3}$,
$20rn<20rn+(56m+38)r=(s+(q+1)(3m+2))(q+1)\leq x(q+1)$ $\leq
[s+(q+1)(5m+1)](q+1)=21rn-(46m+6)<21rn$;  If $x\in I_{4}$,
$21rn<21rn+(28m+26)r=(s+(q+1)(5m+2))(q+1)\leq x(q+1)$ $\leq
[s+(q+1)(7m+2)](q+1)=22rn-(62m+27)r<22rn$;. Hence, there is no skew
symmetric cyclotomic  coset in the defining set $T$. Using the same
method, one can testy that any two cyclotomic  cosets do not form a
skew asymmetric pair. Therefore, if $q$ is the form $74m+31$, and
$0\leq \delta\leq 7m+2$, then $\mathcal{C}$$^{\perp_{h}}$ $\subseteq
\mathcal{C}$. And for $q$ is the form $74m+43$, using the same
method, one also can testy that there is skew symmetric cyclotomic
coset and any two  cyclotomic cosets do not form a skew asymmetric
pair in the defining set $T$. Therefore, $\mathcal{C}$$^{\perp_{h}}$
$\subseteq \mathcal{C}$.

\noindent{\bf  Theory 3.12:} If $q$ is an odd prime power of the
form $74m+31$ or $74m+43$,
 then there exists a q-ary $[[\frac{q^{2}+1}{37},\frac{q^{2}+1}{37}-2d+2,d]]$-quantum MDS
codes, where $2\leq d \leq 14m+6$ is even for $q$ with the form of
$74m+31$;
 and $2\leq d \leq 14m+8$ is even for $q$ with the form of $74m+43$.

\noindent{\it Proof.} Fix $s=\frac{q^{2}+1}{2}$. Consider the
$\eta$-constacyclic codes over $F_{q^{2}}$ of length
$n=\frac{q^{2}+1}{37}$ with defining set
$T=\bigcup_{i=0}^{\delta}C_{s+(q+1)i}$, where $0\leq \delta\leq
7m+2$ for $q$ with the form of $74m+31$, and $0\leq \delta\leq 7m+3$
for $q$ with the form of $74m+43$. By Lemma 3.11, there is
$\mathcal{C}$$^{\perp_{h}}$ $\subseteq \mathcal{C}$. Form lemma 3.1,
we can obtain that $T$ consists of $2\delta+1$ integers
$\{s-(q+1)\delta,\cdots,s-(q+1),s,s+(q+1),\cdots,s+(q+1)\delta\}$.
It implies that $\mathcal{C}$ has minimum distance at least
$2\delta+2$ and the minimum distance $d_{C}$ is even. Hence,
$\mathcal{C}$ is a $q^{2}$-ary $\eta$ constacyclic code with
parameters $[n,n-(2\delta-1),\geq 2\delta+2]$. Combining the
Hermitian construction with quantum Singleton bound, we can obtain a
quantum MDS code with parameters $[[n,n-4\delta-2,2\delta+2]]_{q}$.
Hence, we can obtian $q$
ary-$[[\frac{q^{2}+1}{37},\frac{q^{2}+1}{37}-2d+2,d]]$ quantum MDS
code, where $2\leq d \leq 14m+6$ is even for $q$ with the form of
$74m+31$ and $2\leq d \leq 14m+8$ is even for  $q$ is the form of
$74m+43$.

\begin{center}
Table 5 Quantum MDS codes with length $n=\frac{q^{2}+1}{37}$\\
\begin{tabular}{llllllllllll}
  \hline
  % after \\: \hline or \cline{col1-col2} \cline{col3-col4} ...
  Quantum MDS code            &the form of q        &m  &q &r     &&d  \\
\hline
  $[[866,868-2d,d]]_{179}$   &74m+31              & 2   &179 &180   &&$2\leq d\leq 34$ is even \\

 $[[50,52-2d,d]]_{43}$        &74m+43              &0   &43 &44   &&$2\leq d\leq 8$  is even \\

  \hline
  \end{tabular}
\end{center}

\subsection{ New  Quantum MDS Codes of Lenght
$n=\frac{q^{2}+1}{41}$}

Let $a=41$ and $q$ be an odd prime power with the form $82m+9$ or
$82m+73$, where $m$ is a positive integer. Then $41$ is a divisor of
$q^{2}+1$. Let $n=\frac{q^{2}+1}{41}$, $r=q+1$ and $\eta\in
F_{q^{2}}$ be a primitive $r$th root of unity.  We consider
$\eta$-constacyclic codes over $F_{q^{2}}$ of length $n$ to
construct quantum codes. First, we give a sufficient condition for
$\eta$-constacyclic codes over $F_{q^{2}}$ of length $n$ which
contain their Hermitian duals.

\noindent{\bf  Lemma 3.13:} Let $q$ be an odd prime power,
$n=\frac{q^{2}+1}{41}>q$. Suppose $\mathcal{C}$ is a $q^{2}$-ary
constacyclic code of length $n$  with define  set
$T=\bigcup_{i=0}^{\delta}C_{s+(q+1)i}$. For $m\geq 0$ is a positive
integer, then

(1)  if $q$ is the form $82m+9$,  and  $0\leq \delta\leq 9m$, then
$\mathcal{C}$$^{\perp_{h}}$ $\subseteq \mathcal{C}$.

(2) if $q$ is the form $82m+73$, and  $0\leq \delta\leq 9m+7$, then
$\mathcal{C}$$^{\perp_{h}}$ $\subseteq \mathcal{C}$.

\noindent{\it Proof.} According to Lemma 2.2 and Lemma 2.3, one
obtain that $\mathcal{C}$$^{\perp_{h}}$ $\subseteq \mathcal{C}$ if
and only if there is no  skew symmetric cyclotomic  coset and any
two  cyclotomic  cosets do not form a skew asymmetric pair in the
defining set $T$. For $q$ is the form $82m+9$,
$n=\frac{q^{2}+1}{41}=164m^{2}+36m+2$,
$s=\frac{q^{2}+1}{2}=20n+\frac{n}{2}$. Let $x,y \in
\Omega=\{s+(q+1)j|0\leq j\leq 9m\}$. one only testify that $x+yq
\not\equiv 0$ mod $rn$ holds. To do this, we divide $\Omega$ into
five parts such that $\Omega=\bigcup_{i=1}^{4} I_{i}$, where
$I_{1}=[s,s+(q+1)m]$, $I_{2}=[s+(q+1)(m+1),s+(q+1)3m]$,
$I_{3}=[s+(q+1)(3m+1),s+(q+1)5m]$,
$I_{4}=[s+(q+1)(5m+1),s+(q+1)(7m+1)]$ and
$I_{5}=[s+(q+1)(7m+2),s+(q+1)(9m)]$. First, we testify that any
cyclotomic  coset is not skew symmetric. If $x\in I_{1}$,
$20rn<\frac{(q^{2}+1)(q+1)}{82}+20rn=s(q+1)\leq x(q+1)$ $\leq
[s+(q+1)m](q+1)= 20rn+(164m^{2}+28m+1)r<21rn$; If $x\in I_{2}$,
$21rn<21rn+(74m+9)r=(s+(q+1)(m+1))(q+1)\leq x(q+1)$ $\leq
[s+(q+1)(3m+1)](q+1)\leq 22rn-(24m+3)r<22rn$;
 If $x\in I_{3}$,
$22rn<22rn+(58m+7)r=(s+(q+1)(3m+1))(q+1)\leq x(q+1)$ $\leq
[s+(q+1)5m](q+1)=23rn-(40m+5)<23rn$;  If $x\in I_{4}$,
$23rn<23rn+(42m+5)r=(s+(q+1)(5m+1))(q+1)\leq x(q+1)$ $\leq
[s+(q+1)7m](q+1)=24rn-(68m+1)r<24rn$; If $x\in I_{5}$,
$24rn<24rn+(26m+5)r=(s+(q+1)(7m+1))(q+1)\leq x(q+1)$ $\leq
[s+(q+1)9m](q+1)=25rn-(72m+9)r<25rn$;. Hence, there is no skew
symmetric cyclotomic  coset in the defining set $T$. Using the same
method, one can testy that any two cyclotomic  cosets do not form a
skew asymmetric pair. Therefore, if $q$ is the form $82m+9$, and
$0\leq \delta\leq 9m$, then $\mathcal{C}$$^{\perp_{h}}$ $\subseteq
\mathcal{C}$. And for $q$ is the form $82m+73$, using the same
method, one also can testy that there is skew symmetric cyclotomic
coset and any two  cyclotomic cosets do not form a skew asymmetric
pair in the defining set $T$. Therefore, $\mathcal{C}$$^{\perp_{h}}$
$\subseteq \mathcal{C}$.

\noindent{\bf  Theory 3.14:} If $q$ is an odd prime power of the
form $82m+9$ or $82m+73$,
 then there exists a q-ary $[[\frac{q^{2}+1}{41},\frac{q^{2}+1}{41}-2d+2,d]]$-quantum MDS
codes, where $2\leq d \leq 18m+2$ is even for $q$ with the form of
$82m+9$;
 and $2\leq d \leq 18m+16$ is even for $q$ with the form of $82m+73$.

\noindent{\it Proof.} Fix $s=\frac{q^{2}+1}{2}$. Consider the
$\eta$-constacyclic codes over $F_{q^{2}}$ of length
$n=\frac{q^{2}+1}{41}$ with defining set
$T=\bigcup_{i=0}^{\delta}C_{s+(q+1)i}$, where $0\leq \delta\leq 9m$
for $q$ with the form of $82m+9$, and $0\leq \delta\leq 9m+7$ for
$q$ with the form of $82m+73$. By Lemma 3.13, there is
$\mathcal{C}$$^{\perp_{h}}$ $\subseteq \mathcal{C}$. Form lemma 3.1,
we can obtain that $T$ consists of $2\delta+1$ integers
$\{s-(q+1)\delta,\cdots,s-(q+1),s,s+(q+1),\cdots,s+(q+1)\delta\}$.
It implies that $\mathcal{C}$ has minimum distance at least
$2\delta+2$ and the minimum distance $d_{C}$ is even. Hence,
$\mathcal{C}$ is a $q^{2}$-ary $\eta$ constacyclic code with
parameters $[n,n-(2\delta-1),\geq 2\delta+2]$. Combining the
Hermitian construction with quantum Singleton bound, we can obtain a
quantum MDS code with parameters $[[n,n-4\delta-2,2\delta+2]]_{q}$.
Hence, we can obtian $q$
ary-$[[\frac{q^{2}+1}{41},\frac{q^{2}+1}{41}-2d+2,d]]$ quantum MDS
code, where $2\leq d \leq 18m+2$ is even for $q$ with the form of
$82m+9$ and $2\leq d \leq 18m+16$ is even for  $q$ is the form of
$82m+73$.

\begin{center}
Table 6 Quantum MDS codes with length $n=\frac{q^{2}+1}{41}$\\
\begin{tabular}{llllllllll}
  \hline
  % after \\: \hline or \cline{col1-col2} \cline{col3-col4} ...
  Quantum MDS code            &the form of q        &m &q  &r     &&d  \\
\hline
  $[[730,732-2d,d]]_{173}$    &82m+9                &2 &173  &174   &&$2\leq d\leq 38$ is even \\

 $[[130,132-2d,d]]_{73}$        &82m+73             &0  &73 &74   &&$2\leq d\leq 16$  is even \\

  \hline
  \end{tabular}
\end{center}

\section{ SUMMARY}

In this paper, using constacyclic codes and Hermitain construction,
we construct some new quantum MDS codes of the form $q=2am+t$,
$n=\frac{q^{2}+1}{a}$, where $a=13,17,25,29,37,41$ and
$a|(t^{2}+1)$. In Table 7, we list the quantum MDS constructed in
this paper. All most of these quantum MDS codes are new in the sense
that their parameters are not covered be the codes available in the
literature.\\

\begin{center}
Table 7 New parameters of Quantum MDS codes with length $n=\frac{q^{2}+1}{a}$\\
\begin{tabular}{lllllllllllllll}
  \hline
  % after \\: \hline or \cline{col1-col2} \cline{col3-col4} ...
  Class            &q        & length                &Distance  \\
\hline
  1                &26m+5    &  $\frac{q^{2}+1}{13}$   &$2\leq d\leq 10m+2$ is even \\

                   &26m+21    &  $\frac{q^{2}+1}{13}$   &$2\leq d\leq 10m+8$ is even \\

 2                &34m+13    &  $\frac{q^{2}+1}{17}$   &$2\leq d\leq 10m+4$ is even \\

                  &34m+21    &  $\frac{q^{2}+1}{17}$   &$2\leq d\leq 10m+6$ is even \\

 3                &50m+7    &  $\frac{q^{2}+1}{25}$   &$2\leq d\leq 14m+2$ is even \\

                  &50m+43    &  $\frac{q^{2}+1}{25}$   &$2\leq d\leq 14m+12$ is even \\

 4                &58m+17    &  $\frac{q^{2}+1}{29}$   &$2\leq d\leq 14m+4$ is even \\

                 &58m+41    &  $\frac{q^{2}+1}{29}$   &$2\leq d\leq 14m+10$ is even \\

 5                &74m+31    &  $\frac{q^{2}+1}{37}$   &$2\leq d\leq 14m+6$ is even \\

                 &74m+43    &  $\frac{q^{2}+1}{37}$   &$2\leq d\leq 14m+8$ is even \\

 6                &82m+9    &  $\frac{q^{2}+1}{41}$   &$2\leq d\leq 18m+2$ is even \\

                 &82m+73    &  $\frac{q^{2}+1}{41}$   &$2\leq d\leq 18m+16$ is even \\

  \hline
  \end{tabular}
\end{center}

\section*{Acknowledgment}
This work is supported by the National Key R\&D Program of China
under Grant No. 2017YFB0802400, the National Natural Science
Foundation of China under grant No. 61373171, the National Natural
Science Foundation of China under Grant No.11471011, the Natural
Science Foundation of Shaanxi under Grant No.2017JQ1032 and 111
Project under grant No.B08038.

\bibliographystyle{amsplain}

\end{document}